\documentclass[fleqn,twoside]{article}
\usepackage{espcrc2}

\usepackage{graphicx}
\usepackage[figuresright]{rotating}

\def\be{\begin{equation}}
\def\ee{\end{equation}}

\newcommand{\rpv}{\mbox{$\not\hspace{-0.10cm}R_p$}}
\newcommand{\LV}{\mbox{$\not \hspace{-0.09cm} L \hspace{0.05cm}$}}
\newcommand{\BV}{\mbox{$\not \hspace{-0.09cm} B \hspace{0.05cm}$}}
\newcommand{\gsim}{\lower.7ex\hbox{$\;\stackrel{\textstyle>}{\sim}\;$}}
\newcommand{\lsim}{\lower.7ex\hbox{$\;\stackrel{\textstyle<}{\sim}\;$}}

\newcommand{\AmS}{{\protect\the\textfont2
  A\kern-.1667em\lower.5ex\hbox{M}\kern-.125emS}}

\title{ Exotic SUSY scenarios
       }

\author{R.M. Godbole\address[MCSD]{CERN, Theory Division, CH-1211 Geneva 23, 
        Switzerland}  
        \thanks{Permanent address: Centre for Theoretical Studies,
                Indian Institute of Science, Bangalore, 560 012,
                India.}}
       
\begin{document}

\begin{abstract}
In this talk I discuss some scenarios which involve small extensions of the
ideas usually considered  in the Minimal Supersymmetric Standard Model (MSSM).
I present results of a study of the implication of non-universal gaugino masses 
(NGM) for the invisible decays of the lightest scalar and the  correlation 
of the same with the relic density of the lightest supersymmetric particle 
(LSP) in the Universe.  Further I discuss SUSY with \rpv. Decay of 
$\tilde \chi_1^0$ in the case of dominant \rpv\ and \LV, trilinear
$\lambda$ and $\lambda'$ couplings, and also that of $\tilde \chi_1^0, 
\tilde \chi_1^+$ caused by  the bilinear \rpv, \LV\ couplings are discussed. 
The effect of the latter for the trilepton signal at the Tevatron is presented. 
I end with a discussion of signals of Heavy Majorana Neutrinos (HMN) 
at the LHC.
\vspace{1pc}
\end{abstract}

\setcounter{page}{0}
\thispagestyle{empty}
\begin{flushright}
                                                   hep-ph/0210196  \\
                                                    CERN-TH/2002-268\\
\end{flushright}

\vskip 25pt

\begin{center}

{\bf Exotic SUSY scenarios 
          \footnote{Talk presented at the  International  Conference of High 
Energy Physics, July 24-31,2002, Amsterdam, The Netherlands.}}
\vskip 25pt
  R.M. Godbole\footnote{Permanent address: Centre for Theoretical Studies,
Indian Institute of Science, Bangalore, 560 012, India.}\\
  CERN, Theory Division, CH-1211 Geneva 23, Switzerland\\

\bigskip
           Abstract
\end{center}

\begin{quotation}
\noindent
In this talk I discuss some scenarios which involve small extensions of the
ideas usually considered  in the Minimal Supersymmetric Standard Model (MSSM).
I present results of a study of the implication of non-universal gaugino masses
(NGM) for the invisible decays of the lightest scalar and the  correlation
of the same with the relic density of the lightest supersymmetric particle
(LSP) in the Universe.  Further I discuss SUSY with \rpv. Decay of
$\tilde \chi_1^0$ in the case of dominant \rpv\ and \LV, trilinear
$\lambda$ and $\lambda'$ couplings, and also that of $\tilde \chi_1^0,
\tilde \chi_1^+$ caused by  the bilinear \rpv, \LV\ couplings are discussed.
The effect of the latter for the trilepton signal at the Tevatron is presented.
I end with a discussion of signals of Heavy Majorana Neutrinos (HMN)
at the LHC.
\end{quotation}

\newpage

\maketitle
\section{Introduction}
In this talk I summarize salient features of the results of  four
different investigations. Small departures from the standard assumptions of the
SM and the MSSM are  the common feature unifying all of them. The results 
presented  are based on four abstracts~\cite{abs1,abs2,abs3,abs4} submitted to 
this meeting. 
Supersymmetry and models for non-zero $\nu$ masses indeed form a very big part 
of all the current Beyond the Standard Model (BSM) discussions.  In this talk 
I refer to the effects of relaxing two  of the assumptions normally made: that 
of universal gaugino masses~\cite{abs1} and $R_p$ 
conservation~\cite{abs2,abs3}. 
\rpv\ SUSY provides one of the most economical ways of generating non-zero 
$\nu$ masses.  In the last abstract~\cite{abs4} a new aspect of the collider 
signatures of the heavy Majorana neutrino, an important ingredient of all the
models of $\nu$ masses not involving \rpv\ SUSY, is presented. 

\section{Non-Universal Gaugino Masses}
\begin{figure}[tbh]
\includegraphics*[scale=0.30]{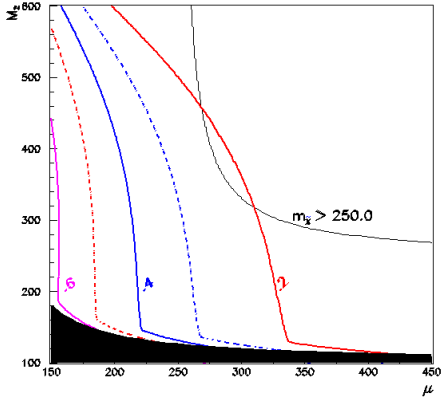}
\vspace{-1.0cm}
\caption{
Contours of  B.R.$(h \rightarrow  \tilde \chi_1^0 \tilde \chi_1^0)$ 
for $r=0.1$ along with the LEP constraints on the
$\tilde\chi_1^+$ mass indicated by the black region. $\tan \beta = 5$
and $m_h = 125$ GeV. 
}
\vspace{-.5cm}
\label{fig1}
\end{figure}

MSSM assumes universal masses  for the $SU(3), SU(2)$ and $U(1)$ gauginos
at the high scale.  This implies $ M_1 \simeq 0.5 M_2 $ at the electroweak
scale, where $M_1, M_2$ are the $U(1), SU(2)$ gaugino masses. However, 
this assumption need not be true even in the very restrictive mSUGRA model 
wherein the non-universality is possible for a non-minimal kinetic term 
for the gauge superfields.  Non-universal gaugino masses are expected
also in models with anomaly-mediated SUSY breaking (AMSB)
or moduli-dominated SUSY breaking.  In general, therefore, we can expect
$M_1 = r M_2$  with $r \ne 0.5$ at the EW scale. We studied the effect of 
such a scenario on the `invisible' decays of the $h$.  A ratio $r$ between 
the two gaugino masses at the EW scale needs
\be
M_1 = 2 r M_2
\label{m1m2}
\ee
at the GUT scale. Most of the above-mentioned models normally imply values
of $r > 1$.  Taking a  phenomenological approach, however,  we consider two 
values of $r < 1$, viz. $0.1,0.2$. As a result of such a non-universality,
for a given $\tilde\chi_1^\pm$ mass, the mass of the $\tilde \chi_1^0$ 
is smaller 
than that in the universal case.  Hence it is possible to have
$h \rightarrow \tilde \chi_1^0 \tilde \chi_1^0$ while respecting all the LEP
constraints, in spite of the theoretical upper limit of $\sim 130 $ GeV on the
mass of the lightest $h$ among the SUSY higgses. Of course it is necessary
to rework~\cite{belbour} all the LEP constraints for the non-universal case.
B.R. ($h \rightarrow \tilde \chi_1^0 \tilde \chi_1^0)$ is 
maximized for moderate $\tan \beta$, small $\mu$ and $M_2$. 
Unlike the case of universal gaugino masses,  
we found large  regions of the $M_2$--$\mu$ plane
where B.R. for the `invisible', $\tilde \chi_1^0 \tilde \chi_1^0$ decay
mode can be as large as $0.65$ {\it even} after all the LEP
constraints are imposed, as shown in Figure~\ref{fig1}.
Such a large `invisible' branching ratio for the $h$ decreases that 
into the $\gamma \gamma$ and $b \bar b$ channels. The  latter  indeed 
provide the best possible signature for the $h$ produced `inclusively' 
and in association with a $W/Z/t\bar t$. If we define, 
$$
R_{\gamma \gamma} = {{\rm B.R. }(h \rightarrow \gamma \gamma)_{SUSY} \over
{\rm B.R.}  (h \rightarrow \gamma \gamma)_{SM}}
$$
and $R_{b \bar b}$ similarly, we find that
$$
R_{\gamma \gamma} = R_{b \bar b} =
1.0 - {\rm B.R. } (h \rightarrow \tilde \chi_1^0 \tilde \chi_1^0).
$$
Thus B.R.$(h \rightarrow \tilde \chi_1^0 \tilde \chi_1^0) \sim 0.3$--$0.4$ 
can mean loss of the signal for the lightest Higgs at the LHC.

A light $\tilde\chi_1^0$ also has implications for the relic density of the
neutralinos in the Universe, as the latter is decided by 
$\sigma (\tilde \chi_1^0 \tilde \chi_1^0  \rightarrow  f^+ f^-)$. For a light 
$\tilde \chi_1^0$, the $Z/h$-mediated $s$ channel process contributes to the 
annihilation. If the $\tilde l_R$  is light  the cross-section also receives a
contribution from the $t$-channel $\tilde l_R$ exchange. 
There is a clear correlation between the expected `invisible' branching 
ratio for the $h$ and the relic density of the $\tilde \chi_1^0$ as the same
couplings are involved. 
\begin{figure}[tbh]
\includegraphics[scale=0.30]{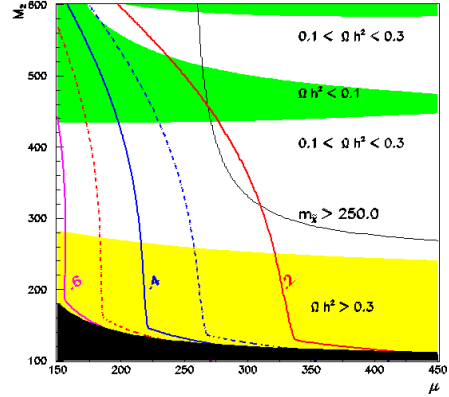}
\vspace{-1cm}
\caption{
Contours of  $B.R. (h \rightarrow  \tilde \chi_1^0 \tilde \chi_1^0)$
for $r=0.1$ along with the DM and  LEP constraints. The white
region corresponds to $0.1 < \Omega h^2 < 0.3$, for $m_0 = 94 $ GeV, 
$\tan \beta = 5$ and $m_h = 125$ GeV. The black region is the LEP-excluded
region.  The lightly (heavily) shaded region corresponds to 
$\Omega h^2 > 0.3~(< 0.1)$.
}
\vspace{-.5cm}
\label{fig2}
\end{figure}
Figure~\ref{fig2} shows results obtained for $r = 0.1$ with 
$\tan \beta = 5$ and $m_{\tilde l_R} \sim 100$ GeV~\cite{abs1} using the 
micrOMEGAs~\cite{fbp} program to calculate the relic density.
These show  that the requirement of an acceptable relic density does 
constrain the $M_2$--$\mu$ plane quite substantially. However, 
there exist large regions of this plane where  the `invisible' branching 
ratio of $h$ is as large as $0.5$--$0.6$, even for `large' $(\sim 200$ GeV) 
$\tilde l_R$,  consistent with the LEP constraints and with an acceptable 
relic density.  The small mass of the $\tilde \chi_1^0$ in this case means that 
the trilepton signal at the Tevatron will be qualitatively different 
from the universal case. Thus this scenario 
can be tested at the Tevatron via the trilepton events.  As a matter of fact,  
one can also obtain a model-independent limit on $M_1$ and hence on  
the $\tilde \chi_1^0$  mass, by considerations of relic density.

\section{Decaying $\tilde \chi_1^0$ in $R_p$-violating theories:}
$B,L$ are symmetries of the SM but not of the MSSM; i.e. it is possible 
to write terms in the Lagrangian which are \LV, \BV\ but respect gauge 
invariance as well as supersymmetry.
Non-zero $\nu$ masses can be generated in \rpv\ supersymmetric theories,
without introducing any new fields; at the tree level via the bilinear
\rpv\ terms in the superpotential and quantum one or two loop level 
via the trilinear $\lambda, \lambda'$ couplings~\cite{borzu1}. A large number 
of these couplings are constrained by low energy processes, cosmological 
arguments as well as explicit collider searches~\cite{limits}. Some of the 
strongest constraints on the $\Delta L =1$ processes come from $\nu$ masses.
As said earlier, since the heavy scale involved in the mass-generation 
mechanism is decided by the sparticle masses, this leads to constraints on the 
size of the \rpv\ couplings and hence to clear predictions of the \LV , \BV\ 
signals in collider processes involving these sparticles. Since some of the
constraints do depend on the details of the specific SUSY model, it is 
important to study the direct effect of the {\it very same} couplings that are 
so indirectly constrained, at the colliders.  Such studies can help
clarify the picture for model building.

The main constraints on  \LV\ couplings $\lambda, \lambda'$  with more than 
one third-generation index come from models for $\nu$  masses and not from 
collider experiments. Probes of these \LV\ couplings at the colliders 
involve studying the physics of third-generation fermions.  The third 
generation sfermions are also likely to give rise to larger virtual effects, 
as they are  expected to be lighter than those of the first two generations.
For $\tilde \chi_1^0, \tilde \chi_1^+$ with masses
of interest at the LHC and NLC, final states with third-generation fermions
including $t$ are possible.  One thus needs a study of the \rpv\ decays of
$\tilde \chi_1^0, \tilde \chi_1^\pm$ retaining effects of the mass of the
third generation fermions, for $L$-violating coupling. An interesting example
is the resonant or non-resonant production  of the $\tilde \tau$ via the
\LV\, $\lambda'_{333}$ coupling in $p p \rightarrow t \bar b \tilde \tau$, 
similar to the case of $t \bar b H^+$ production via the Yukawa coupling.
Even for $\lambda'_{333}$ as small as $0.01$ and $m_{\tilde t} > m_t$,
the rates are appreciable. The $\tilde \tau$ thus produced will have both
$R_p$-conserving decay $\tilde \tau \rightarrow \tau \tilde \chi_1^0, 
\tilde \tau \rightarrow \nu_\tau \tilde \chi_1^-$  for $m_t > m_{\tilde\tau}$ 
and the  $R_p$-violating one  $\tilde \tau \rightarrow b \bar t$
for $m_t < m_{\tilde \tau}$. The final state composition will 
depend on the \rpv\ decays of the $\tilde \chi_1^0$  as well.
Production of $\tilde \tau$ through \rpv\ couplings and its decay
via the {\it same} will give rise to $p p \rightarrow t \bar b \tilde 
\tau X \rightarrow t \bar b t \bar b X$, which  is the same as the expected 
final state for the $H^\pm$ production. Thus the $\tilde \tau$ in this case 
can fake the $H^\pm$ signal.  $R_p$ conserving decays of the $\tilde \tau$ 
and \rpv\ decays of the $\tilde \chi_1^0$ can also produce
\begin{eqnarray}
p p \rightarrow t \bar b \tilde \tau X &\rightarrow (2t) (2b) (2\tau) X\nonumber\\
&\rightarrow tb~(2 \bar b)~\tau \nu_\tau X.
\label{staudec}
\end{eqnarray}
In the latter case the final state has the tell-tale signature of 
\LV, the like-sign fermion pairs. 

We calculated~\cite{abs2} the three-body decays of the LSP for dominant 
\LV\ trilinear couplings and obtained explicit expressions for the most 
general case of complex mass matrices, including the effect of the 
mass of the third-generation fermions. Of course the results depend on 
the $L$--$R$ mixing in the sfermion sector, gaugino--higgsino mixing and 
$\tan \beta$, as well as on the generation structure of the particular 
\LV\ couplings that is large. One can have the `massive' and `massless'  
modes with/without a $t$ in the final state.
\begin{figure*}[tbh]
\includegraphics*[scale=0.35]{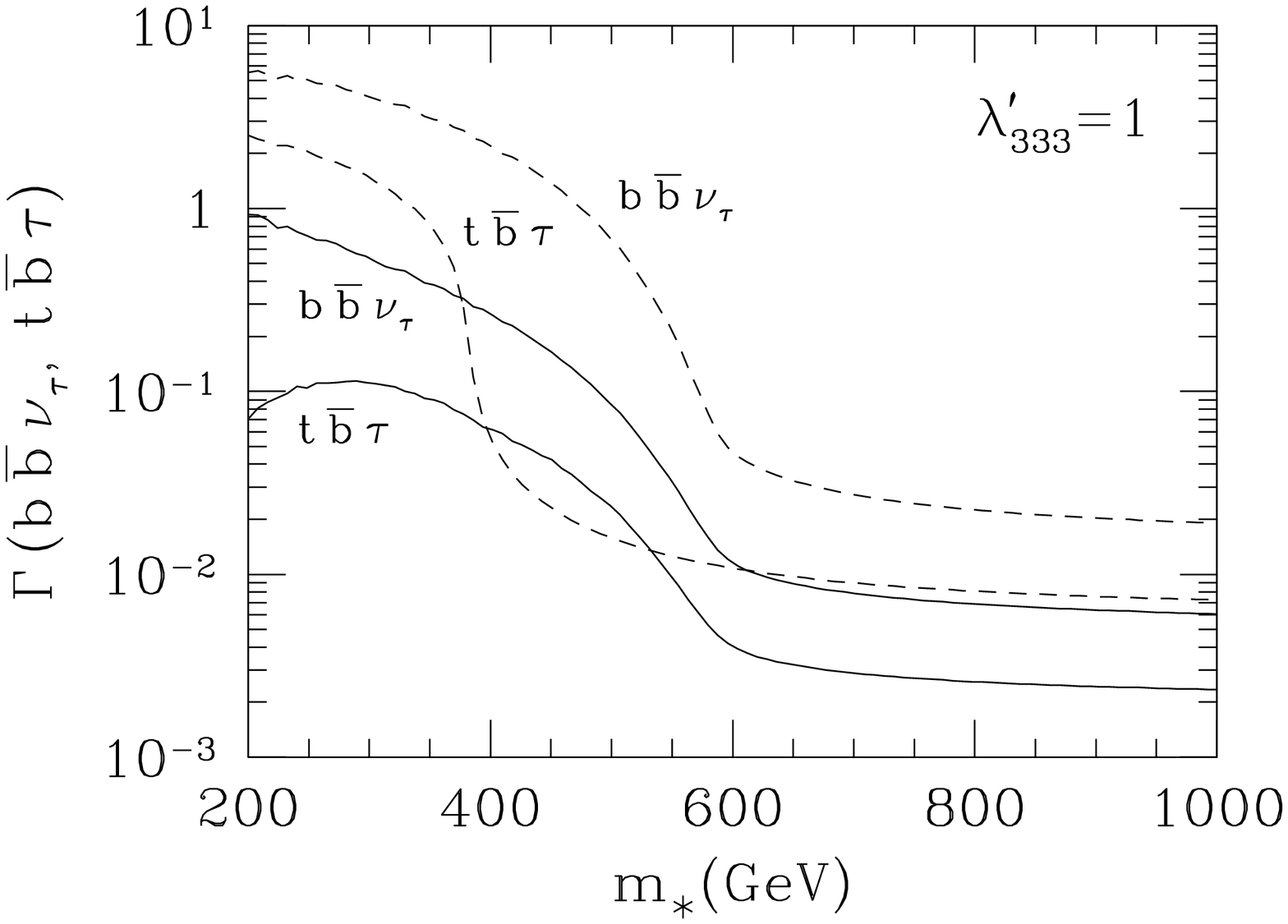}
\includegraphics*[scale=0.35]{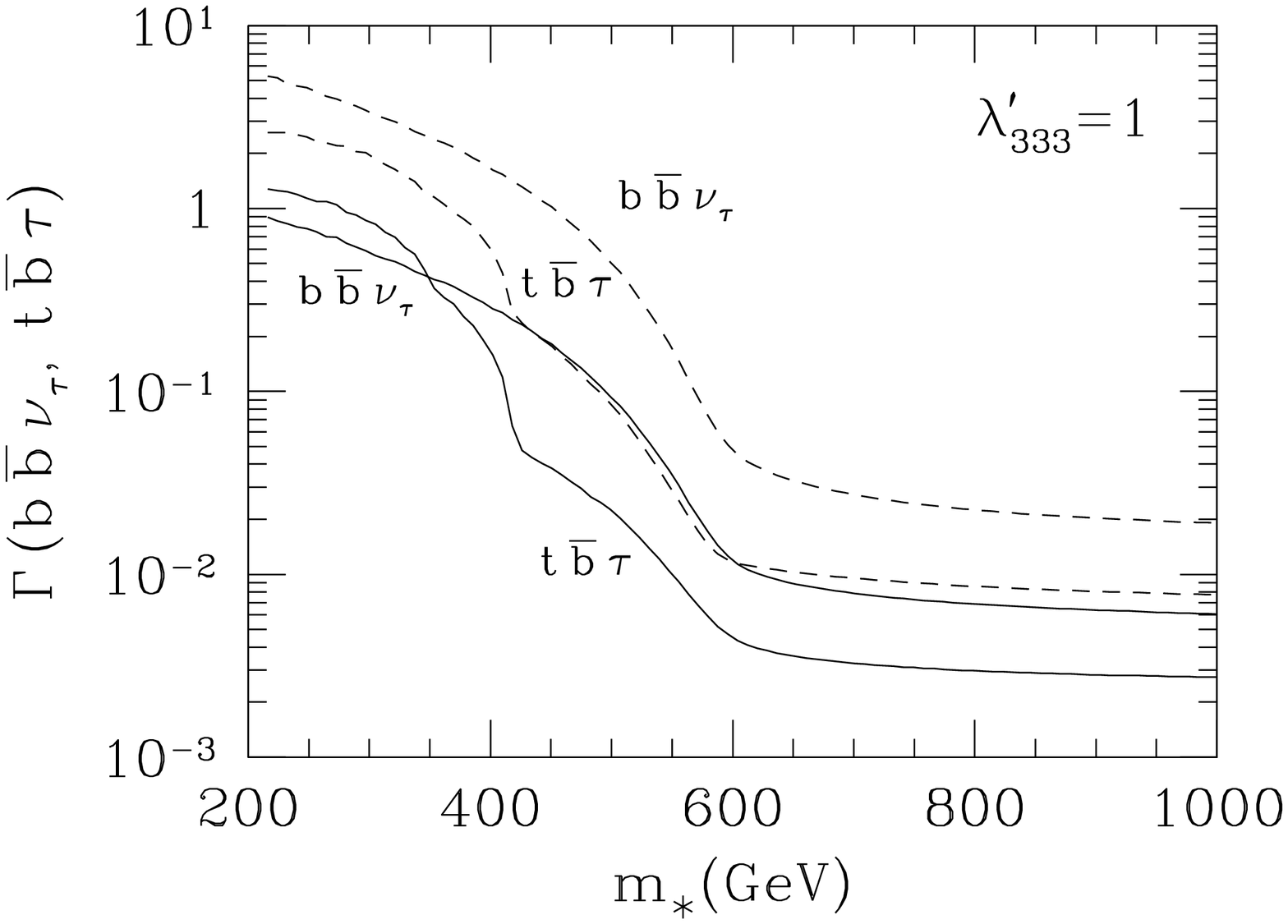}
\vspace{-0.5cm}
\caption{$\tilde \chi_1^0$ widths for no (moderate) $L$--$R$ squark mixing in
left (right) panel. Dashed (solid) lines for a wino (bino) like 
$\tilde \chi_1^0$. $A_t - \mu \cot \beta = 150$ GeV, 
$A_b - \mu \tan \beta = 2000$ GeV for non-zero mixing.}
\vspace{-.5cm}
\label{fig3}
\end{figure*}
Figure~\ref{fig3} shows the results for the case with $\lambda'_{333}$ dominant,
for two different choices of $L$--$R$ mixing in the squark sector and 
different composition of the $\tilde\chi_1^0$. The massive decay mode 
has a large width for low $\tan \beta$ and large higgsino--gaugino mixing.  
Otherwise the massless mode is larger, although the massive mode is 
non-negligible. Even for smaller but dominant \rpv, $\lambda$ and $\lambda'$ 
couplings, with more than one third-generation index, the three-body \rpv\ 
decays of $\tilde \chi_1^0, \tilde \chi_1^+$ can have important 
phenomenological consequences for new particle searches at the future colliders.

In the above discussion the bilinear \rpv\ terms were assumed to be zero. In 
the analysis of Ref.~\cite{abs3}, a particular case of non-zero
$\kappa_3$ is considered.  The \rpv\ and $R_p$-conserving decays 
of the $\tilde\chi_1^0,\tilde\chi_1^\pm$ in this case are given by
Eqs. (\ref{rpvd}) and (\ref{rpcd}) respectively.
\begin{eqnarray}
\tilde \chi_1^0 &\rightarrow \nu_\tau Z^* \rightarrow \nu_\tau f \bar f,~~
\tilde \chi_1^0 \rightarrow \tau W^\pm \rightarrow \tau f \bar f' \nonumber\\
\tilde \chi_1^\pm &\rightarrow \tau Z^* \rightarrow \tau f \bar f,~~
\tilde \chi_1^\pm \rightarrow \nu_\tau W^\pm \rightarrow \nu_\tau  f \bar f',
\label{rpvd}
\end{eqnarray}
and 
\be
\tilde \chi_2^0 \rightarrow \tilde \chi_1^0 \ell^+ \ell^-,
\tilde \chi_1^\pm \rightarrow \tilde  \chi_1^0 \ell \nu_\ell.
\label{rpcd}
\ee
It is clear the \rpv\  decays will enhance the number of leptons in the
final state and give very clear signals. These authors analysed, in an 
mSUGRA picture, the multilepton signals at the Tevatron.  The multilepton 
signal is one of the promising channel for discovery of SUSY at the Tevatron. 
Hence such an analysis is important. This one shows that the reach in the 
multilepton channel for the \rpv\ case is much better than for the 
$R_p$-conserving case at large values of $m_0$.  For small $m_0$ 
the situation is reversed, owing to slepton-mediated $\tilde \chi_2^0$ decay.
\section{Signals for Heavy Majorana Neutrino}
The authors~\cite{abs4} consider models of $\nu$ masses which have
an isosinglet neutrino $N$ that mixes with the ordinary light one but wherein 
the mass and the mixing of the $N$ are not linked together,  unlike in
the usual models based on  he see-saw mechanism.  This mixing is  then
treated as a phenomenological parameter constrained by the LEP data.
It is possible in these models  to have $N$ mass  `naturally'
in the range: $100 < M_N < 1000$ GeV~\cite{gluza}. The issue
of characteristic \LV\ signals for the HMN via like sign dilepton 
(LSD) events.  was revisited in this investigation. 
They considered both the fusion contribution from
\be
q_i + \bar q_j \rightarrow q_k + \bar q_l + W^{+*} + W ^{+*} \rightarrow 
q_k + \bar q_l + l^+ + l ^+
\ee
where the $N$ is exchanged in the $t$ channel and also from the
resonant production of $N$ via 
\begin{eqnarray}
q_i + \bar q_j \rightarrow W^{+*} \rightarrow l^+ N &\rightarrow l^+ +  l^+  + 
W^-\nonumber\\
&\rightarrow l^+ l^+ q_k \bar q_l.
\end{eqnarray}
Explicit formulae were obtained in helicity formalism and
errors in the earlier calculations~\cite{almeida}, traced to dropping of the
ghost diagrams, were corrected. The predicted  LSD cross-section increases
by about $20 \%$ as a result of this. The reach of the LHC, just from rates, is
$M_N = 200$--$250$  GeV. Detailed simulations will be needed to make
these conclusions firmer, as the cuts can then be optimized.

\end{document}